\documentclass[conference]{IEEEtran}
\IEEEoverridecommandlockouts

\usepackage{cite}
\usepackage{amsmath,amssymb,amsfonts}
\usepackage{algorithmic}
\usepackage{graphicx}
\usepackage{textcomp}
\usepackage{xcolor}

\usepackage{graphicx}
\usepackage{dcolumn}
\usepackage{bm}

\usepackage{color}
\usepackage{subcaption}

\usepackage[utf8]{inputenc}
\usepackage[T1]{fontenc}
\usepackage{mathptmx}
\usepackage{etoolbox}
\usepackage{comment}

\usepackage[colorlinks = true]{hyperref}

\hypersetup{colorlinks=true, linkcolor=blue, citecolor=blue, urlcolor=black, breaklinks=true}

\graphicspath{{./figure/}}

\def\BibTeX{{\rm B\kern-.05em{\sc i\kern-.025em b}\kern-.08em
    T\kern-.1667em\lower.7ex\hbox{E}\kern-.125emX}}
\begin{document}

\title{Towards efficient and secure quantum-classical communication networks\\
}

\author{Pei Zeng$^{1}$\thanks{$^1$Pritzker School of Molecular Engineering, The University of Chicago, Chicago, IL 60637, USA.},
Debayan Bandyopadhyay$^{1}$,
José A. Méndez Méndez$^{1}$,
Nolan Bitner$^{1,2,3}$\thanks{$^2$Materials Science Division, Argonne National Laboratory, Lemont, IL 60439, USA.}\thanks{$^3$Center for Molecular Engineering, Argonne National Laboratory, Lemont, IL 60439, USA.},
Alexander Kolar$^{1}$,\\
Michael T. Solomon$^{1,2,3}$,
F. Joseph Heremans$^{1,2,3}$,
David D. Awschalom$^{1,2,3,4}$\thanks{$^4$Department of Physics, University of Chicago, Chicago, IL 60637, USA.},
Liang Jiang$^{1}$, and
Junyu Liu$^{1,5,6,7}$\thanks{$^5$Department of Computer Science, The University of Chicago, Chicago, IL 60637, USA.}\thanks{$^6$SeQure, Chicago, IL 60637, USA.}\thanks{$^7$Department of Computer Science, The University of Pittsburgh, Pittsburgh, PA 15260, USA.}\\
Emails: \{junyuliucaltech@gmail.com, liangjiang@uchicago.edu, peizeng@uchicago.edu\}}

\maketitle

\begin{abstract}
The rapid advancement of quantum technologies calls for the design and deployment of quantum-safe cryptographic protocols and communication networks. 
There are two primary approaches to achieving quantum-resistant security: quantum key distribution (QKD) and post-quantum cryptography (PQC). 
While each offers unique advantages, both have drawbacks in practical implementation. 
In this work, we introduce the pros and cons of these protocols and explore how they can be combined to achieve a higher level of security and/or improved performance in key distribution. 
We hope our discussion inspires further research into the design of hybrid cryptographic protocols for quantum-classical communication networks.
\end{abstract}

\begin{IEEEkeywords}
quantum key distribution, post-quantum cryptography, key encapsulation mechanism, quantum network.
\end{IEEEkeywords}

\textbf{Introduction}: The rapid development of quantum technologies across various fields, such as quantum computing~\cite{google2023surfacecode,quera2023quantum,deng2023gaussian}, quantum sensing~\cite{degen2017quantum}, and quantum communication~\cite{wehner2018quantuminternet,xu2020secure}, has garnered significant attention in recent years. Notably, the advancement and scale-up of quantum computers have raised serious concerns regarding the security of current cryptographic systems, particularly Rivest-Shamir-Adleman (RSA) encryption\textemdash a type of asymmetric public-key cryptography\textemdash which is based on the difficulty of factoring large numbers and is one of the primary algorithms securing today's digital communications~\cite{rivest1978method}. Quantum computers leverage the principles of quantum mechanics to solve specific problems that classical computers struggle with, including cryptographic challenges.
Many widely used asymmetric cryptographic algorithms rely on the difficulty of three key mathematical problems: the integer factorization problem, the discrete logarithm problem, and the elliptic-curve discrete logarithm problem.
For instance, the security of RSA encryption depends on the difficulty of factoring a large composite number $N = pq$, where $p$ and $q$ are secret prime numbers~\cite{rivest1978method}. The strength of RSA encryption lies in the computational challenge of decomposing $N$ into its prime factors.  However, this security foundation is at risk. In 1994, Shor introduced a quantum algorithm capable of efficiently factorizing any integer $N$~\cite{shor1994,shor1999polynomial}. The potential for large-scale, fault-tolerant quantum computers to execute Shor’s algorithm poses a significant threat to the integrity of current cryptographic systems.
This necessitates the design of new cryptography systems which are robust against quantum attacks.

One approach to designing quantum-safe cryptographic systems is to base their security on the principles of quantum mechanics~\cite{gisin_quantum_2002,zapatero2023}. Extensive research has been conducted in this area, covering topics such as key  distribution~\cite{bennett1984quantum,ekert1992quantum}, random number generation~\cite{herrero2017quantum}, digital signatures~\cite{gottesman2001quantum}, position-based cryptography~\cite{bluhm2022single}, bit commitment~\cite{lunghi2013experimental}, and oblivious transfer~\cite{crepeau1994quantum}, among others.
Quantum key distribution (QKD), in particular, has been proven to be information-theoretically secure~\cite{zapatero2023} and is implementable with current commercial optical devices. The concept of QKD was first introduced by Bennett and Brassard in 1984~\cite{bennett1984quantum}, and a full security proof was completed at the end of the last century~\cite{lo1999unconditional}. Since then, QKD protocol designs have evolved to enhance their security in practical implementations.
To avoid information leakage due to multi-photon components in standard laser sources, decoy-state methods~\cite{Hwang2003quantum,lo2005decoy,wang2005beating} were proposed in the early 21st century, reducing the need for ideal quantum sources. Additionally, to address vulnerabilities in measurement devices, the measurement-device-independent QKD protocol~\cite{lo2012measurement} was introduced, effectively preventing attacks on detectors.
More recently, twin-field QKD protocols~\cite{lucamarini2018overcoming,ma2018phase,zeng2022mode} have significantly improved long-distance performance, even without relying on relay nodes or quantum repeaters. Furthermore, the demonstration of device-independent QKD~\cite{umesh2019fully,xu2022device} offers a solution with the highest level of security, eliminating the need for trust in quantum devices.

Another approach is to explore alternative computationally hard mathematical problems, which has sparked interest in the field of post-quantum cryptography (PQC) \cite{bernstein2017post}. PQC focuses on designing cryptographic algorithms that are believed to be secure even in the face of potential quantum computer attacks. This field has gained significant attention as researchers seek to develop quantum-resistant classical cryptographic methods that can safeguard data in the post-quantum era. In 2017, the National Institute of Standards and Technology (NIST) launched an initiative to standardize one or more quantum-resistant public-key cryptographic algorithms. After soliciting minimum acceptability and submission requirements, and evaluation criteria, an initial round of 69 candidate algorithms were reviewed for key encapsulation mechanisms (KEM) and digital signature algorithms (DSA). Through subsequent revisions of candidate algorithms in additional evaluation rounds, NIST selected one KEM and three DSA algorithms for implementation in the first official PQC standardization.
On August 12, 2024, NIST finalized its first set of standards for post-quantum cryptography, aimed at protecting data from future quantum-based cyber threats that could potentially compromise current cryptographic systems \cite{NISTFIPS203,NISTFIPS204,NISTFIPS205}. The selected algorithms include CRYSTALS-Kyber \cite{bos2018crystals} for KEM, and CRYSTALS-Dilithium \cite{lyubashevsky2020crystals} and SPHINCS+ \cite{bernstein2015sphincs,bernstein2019sphincs+} for DSA. The CRYSTALS-Kyber and CRYSTALS-Dilithium algorithms are based on lattice-based cryptography, while SPHINCS+ is a hash-based cryptographic algorithm.

\textbf{Hybrid quantum-classical networks}: QKD and PQC each have distinct advantages and disadvantages in terms of efficiency and security. QKD offers unmatched security by leveraging the principles of quantum mechanics, ensuring that any eavesdropping attempt can be detected and thwarted. However, its efficiency is hampered by the need for specialized hardware, quantum communication channels, and its limited range, making it less practical for widespread deployment~\cite{diamanti_practical_2016}. In contrast, PQC is highly efficient, as it can be implemented using existing classical infrastructure and is designed to secure data against future quantum attacks. However, PQC's security relies on the hardness of specific mathematical problems, which, while currently resistant to classical or quantum attacks, could be challenged by future developments in classical or quantum algorithms. Combining QKD and PQC can create a more resilient cryptographic framework: QKD can provide secure key distribution at the physical layer, while PQC ensures data protection at the computational layer. This hybrid approach leverages the strengths of both technologies, enhancing security while maintaining practical efficiency. Existing works include using PQC certification for QKD \cite{wang2021experimental} and joint PQC-QKD protocols, where QKD is used for raw-key transmission while a PQC subsystem transmits parity bits for information reconciliation \cite{djordjevic2020joint}. 

Moreover, in a recent work~\cite{ThisWork}, we propose a hybrid key-distribution framework that integrates QKD with PQC, combining the strengths of both techniques to build more flexible and quantum-resistant communication networks. Our hybrid approach allows for enhanced performance or security, depending on the specific needs of the communication infrastructure. We consider two core key-distribution mechanisms: a PQC-based QEM approach and a point-to-point QKD protocol. KEMs use public-key encryption to securely exchange symmetric keys, while QKD generates symmetric keys via quantum measurements. We explore hybrid network designs that integrate these two mechanisms, addressing their respective shortcomings to improve the robustness of cryptographic systems over long distances. To enhance the performance of key distribution rates in long-distance communication, we propose a series-connection protocol which leverages both QKD and KEM to achieve increased performance in a practical regime where users only have access to limited computational power (for key generation, encryption, and decryption). In this protocol, Alice and Bob distribute key bits with nearby data centers using short-distance QKD devices. The data centers then use KEM to distribute symmetric keys over long distances using  high-performance computers to achieve higher key rates than those possible on individual users' hardware. This approach enables higher key rates than are possible with either individual protocol by taking advantage of QKD's performance at short distances and KEM's performance at long distances. To further enhance security, we propose XOR-based and secret-sharing-based parallel key distribution schemes. The XOR scheme combines keys from both KEM and QKD channels, ensuring that the final key is secure even if one of the channels is compromised. The secret-sharing approach, based on Shamir’s secret-sharing scheme~\cite{Simmons1990how}, distributes secrets across multiple key distribution channels, making it theoretically impossible for an attacker to learn the final key unless they gain access to a critical number of shares. The secret sharing scheme is suitable when two end users share multiple key distribution links, in which case the overall key distribution efficiency can be higher than the XOR-based scheme. Furthermore, by introducing advanced linear-code or multi-linear secret-sharing schemes, this approach allows for designing custom trust hierarchies within the network, enhancing the flexibility and adaptability to different security needs. In addition to these hybrid PQC-QKD designs, we discuss methods for evaluating and optimizing their performance across different network typologies. To quantitatively assess the security of the overall key distribution network, we introduce a systematic method to check the minimal access structure of the whole network. A minimal access structure is a subset of the QKD links, KEM links, and trusted nodes such that if the adversary breaks the elements in one subset, they can reveal the final key information shared by two end users. By identifying the set of all minimal access structures of the network and quantifying the security of different links and nodes, we can check the most vulnerable part in the network and design counter-measures accordingly. Our analysis shows that a hybrid PQC-QKD system offers advantages over using either method alone in many scenarios. The series-connection protocol enhances long-distance performance, making QKD a more viable option for real-world deployment, while the parallel-connection design improves security, offering protection against physical and algorithmic attacks. By integrating PQC and QKD, users can create quantum-safe communication systems tailored to specific performance and security requirements, making this approach adaptable for a wide range of applications, from metropolitan networks to global communication infrastructures.

\textbf{Quantum-classical switch}: Another possible method for combining PQC and QKD involves dynamical switching between PQC and QKD. As discussed before, PQC is efficient, but the complexity theory at its foundation may in the future be challenged by the development of novel quantum or classical algorithms. In fact, there are two famous attempts that have tried to break lattice-based cryptography using quantum algorithms \cite{eldar2016efficient,chen2024quantum}, although these two methods were shown to be incorrect. Additionally, an algorithm using supersingular isogeny key exchange (specifically, SIKE or SIDH) \cite{jao2011towards}, which had been previously considered by NIST, was proven to be insecure by Castryck and Decru (2023) \cite{castryck2023efficient}. As a result, this algorithm was removed from NIST's selected algorithm list in 2022. 

To mitigate future potential hacks from either classical or quantum computers against PQC, an efficient switch system should be developed for customers requiring higher security. In cases where future developments provide evidence indicating vulnerabilities in a given cryptographic system, that system should systematically evaluate potential risks and determine the appropriate switching process, which would encompass both software and hardware adjustments. This necessitates more detailed development for both PQC hardware (classical computing devices) and QKD hardware (quantum devices generating and receiving photons, along with optical fibers or vacuum beam guides \cite{huang2024vacuum}), as well as a clear scheme for when and how to switch. Similar switching schemes should also include the ability to switch from one PQC algorithm to another, adjust key sizes in both QKD and PQC, or switch back from QKD to PQC if the users feel PQC is secure enough and wish to maximize efficiency. For instance, it is generally believed that, under current standards, hash-based algorithms like SPHINCS+ may offer greater security than lattice-based algorithms like CRYSTALS-Dilithium, as the security of hash-based algorithms is purely based on the robustness of hash functions themselves. Therefore, it is crucial to detect potential hacks from classical or quantum computers and evaluate the associated risks. If an algorithm is compromised, methods must be designed to swiftly transition to a more secure alternative.

\textbf{Conclusion}: Quantum resistance is a critical feature for future cryptographic systems, with both classical approaches (PQC) and quantum approaches (QKD) offering viable solutions. However, each has its own advantages and disadvantages. In this article, we propose that these two approaches, while distinct, can be complementary, and that combining them could balance security and efficiency. We highlight two key ideas: first, the design of hybrid quantum-classical networks that integrate both PQC and QKD, and second, the development of a quantum-classical switch capable of transitioning between PQC and QKD in response to hacking attempts or when efficiency becomes a primary concern. This combination of PQC and QKD presents an exciting research frontier, promising more efficient, secure, and private services for future users by harnessing the strengths of both classical cryptography and quantum physics. Such advancements could play a crucial role in the development of quantum data centers \cite{liu2022quantum,liu2024quantum} and the quantum internet \cite{kimble2008quantum}.


\section*{Acknowledgment}
We acknowledge support from the ARO (W911NF-23-1-0077), the ARO MURI (W911NF-21-1-0325), the AFOSR MURI (FA9550-19-1-0399, FA9550-21-1-0209, FA9550-23-1-0330, FA9550-23-1-0338), the DARPA (HR0011-24-9-0359, HR0011-24-9-0361), the NSF (OMA-1936118, ERC-1941583, OMA-2137642, OSI-2326767, CCF-2312755), NTT Research, the Packard Foundation (2020-71479), and the Marshall and Arlene Bennett Family Research Program. Additional support was provided by Q-NEXT, part of the U.S. Department of Energy, Office of Science, National Quantum Information Science Research Centers, and the Advanced Scientific Computing Research (ASCR) program under contract number DE-AC02-06CH11357 as part of the InterQnet quantum networking project. J.L. acknowledges startup funds provided by the University of Pittsburgh and funding from IBM Quantum through the Chicago Quantum Exchange.

\bibliographystyle{ieeetr}
\bibliography{bibPQCQKD.bib}

\begin{thebibliography}{10}

\bibitem{google2023surfacecode}
G.~Q. AI, ``Suppressing quantum errors by scaling a surface code logical qubit,'' {\em Nature}, vol.~605, pp.~669--674, 2023.

\bibitem{quera2023quantum}
Q.~Q. Inc., ``Logical quantum processor based on reconfigurable atom arrays,'' {\em Nature}, vol.~605, pp.~268--272, 2023.

\bibitem{deng2023gaussian}
Y.-H. Deng, Y.-C. Gu, H.-L. Liu, S.-Q. Gong, H.~Su, Z.-J. Zhang, H.-Y. Tang, M.-H. Jia, J.-M. Xu, M.-C. Chen, J.~Qin, L.-C. Peng, J.~Yan, Y.~Hu, J.~Huang, H.~Li, Y.~Li, Y.~Chen, X.~Jiang, L.~Gan, G.~Yang, L.~You, L.~Li, H.-S. Zhong, H.~Wang, N.-L. Liu, J.~J. Renema, C.-Y. Lu, and J.-W. Pan, ``Gaussian boson sampling with pseudo-photon-number-resolving detectors and quantum computational advantage,'' {\em Phys. Rev. Lett.}, vol.~131, p.~150601, Oct 2023.

\bibitem{degen2017quantum}
C.~Degen, F.~Reinhard, and P.~Cappellaro, ``Quantum sensing,'' {\em Reviews of Modern Physics}, vol.~89, no.~3, p.~035002, 2017.

\bibitem{wehner2018quantuminternet}
S.~Wehner, D.~Elkouss, and R.~Hanson, ``Quantum internet: A vision for the road ahead,'' {\em Science}, vol.~362, no.~6412, p.~eaam9288, 2018.

\bibitem{xu2020secure}
F.~Xu, X.~Ma, Q.~Zhang, H.-K. Lo, and J.-W. Pan, ``Secure quantum key distribution with realistic devices,'' {\em Rev. Mod. Phys.}, vol.~92, p.~025002, May 2020.

\bibitem{rivest1978method}
R.~L. Rivest, A.~Shamir, and L.~Adleman, ``A method for obtaining digital signatures and public-key cryptosystems,'' {\em Communications of the ACM}, vol.~21, no.~2, pp.~120--126, 1978.

\bibitem{shor1994}
P.~Shor, ``Algorithms for quantum computation: discrete logarithms and factoring,'' in {\em Proceedings 35th Annual Symposium on Foundations of Computer Science}, pp.~124--134, 1994.

\bibitem{shor1999polynomial}
P.~W. Shor, ``Polynomial-time algorithms for prime factorization and discrete logarithms on a quantum computer,'' {\em SIAM review}, vol.~41, no.~2, pp.~303--332, 1999.

\bibitem{gisin_quantum_2002}
N.~Gisin, G.~Ribordy, W.~Tittel, and H.~Zbinden, ``Quantum cryptography,'' {\em Reviews of Modern Physics}, vol.~74, pp.~145--195, Mar. 2002.
\newblock Publisher: American Physical Society.

\bibitem{zapatero2023}
V.~Zapatero, T.~van Leent, R.~Arnon-Friedman, W.-Z. Liu, Q.~Zhang, H.~Weinfurter, and M.~Curty, ``Advances in device-independent quantum key distribution,'' {\em npj Quantum Information}, vol.~9, no.~1, p.~10, 2023.

\bibitem{bennett1984quantum}
C.~H. Bennett and G.~Brassard, ``Quantum cryptography: Public key distribution and coin tossing,'' in {\em Proceedings of IEEE International Conference on Computers, Systems and Signal Processing}, (Bangalore, India), pp.~175--179, 1984.

\bibitem{ekert1992quantum}
A.~K. Ekert, ``Quantum cryptography and bell’s theorem,'' in {\em Quantum Measurements in Optics}, pp.~413--418, Springer, 1992.

\bibitem{herrero2017quantum}
M.~Herrero-Collantes and J.~C. Garcia-Escartin, ``Quantum random number generators,'' {\em Reviews of Modern Physics}, vol.~89, p.~015004, 2017.

\bibitem{gottesman2001quantum}
D.~Gottesman and I.~L. Chuang, ``Quantum digital signatures,'' {\em arXiv preprint quant-ph/0105032}, 2001.

\bibitem{bluhm2022single}
A.~Bluhm, M.~Christandl, and F.~Speelman, ``A single-qubit position verification protocol that is secure against multi-qubit attacks,'' {\em Nature Physics}, vol.~18, no.~6, pp.~623--626, 2022.

\bibitem{lunghi2013experimental}
T.~Lunghi, J.~Kaniewski, F.~Bussi\`eres, R.~Houlmann, M.~Tomamichel, A.~Kent, N.~Gisin, S.~Wehner, and H.~Zbinden, ``Experimental bit commitment based on quantum communication and special relativity,'' {\em Phys. Rev. Lett.}, vol.~111, p.~180504, Nov 2013.

\bibitem{crepeau1994quantum}
C.~Cr{\'e}peau, ``Quantum oblivious transfer,'' {\em Journal of Modern Optics}, vol.~41, no.~12, pp.~2445--2454, 1994.

\bibitem{lo1999unconditional}
H.-K. Lo and H.~F. Chau, ``Unconditional security of quantum key distribution over arbitrarily long distances,'' {\em science}, vol.~283, no.~5410, pp.~2050--2056, 1999.

\bibitem{Hwang2003quantum}
W.-Y. Hwang, ``Quantum key distribution with high loss: Toward global secure communication,'' {\em Phys. Rev. Lett.}, vol.~91, p.~057901, Aug 2003.

\bibitem{lo2005decoy}
H.-K. Lo, X.~Ma, and K.~Chen, ``Decoy state quantum key distribution,'' {\em Phys. Rev. Lett.}, vol.~94, p.~230504, Jun 2005.

\bibitem{wang2005beating}
X.-B. Wang, ``Beating the photon-number-splitting attack in practical quantum cryptography,'' {\em Phys. Rev. Lett.}, vol.~94, p.~230503, Jun 2005.

\bibitem{lo2012measurement}
H.-K. Lo, M.~Curty, and B.~Qi, ``Measurement-device-independent quantum key distribution,'' {\em Phys. Rev. Lett.}, vol.~108, p.~130503, Mar 2012.

\bibitem{lucamarini2018overcoming}
M.~Lucamarini, Z.~L. Yuan, J.~F. Dynes, and A.~J. Shields, ``Overcoming the rate--distance limit of quantum key distribution without quantum repeaters,'' {\em Nature}, vol.~557, no.~7705, pp.~400--403, 2018.

\bibitem{ma2018phase}
X.~Ma, P.~Zeng, and H.~Zhou, ``Phase-matching quantum key distribution,'' {\em Phys. Rev. X}, vol.~8, p.~031043, Aug 2018.

\bibitem{zeng2022mode}
P.~Zeng, H.~Zhou, W.~Wu, and X.~Ma, ``Mode-pairing quantum key distribution,'' {\em Nature Communications}, vol.~13, no.~1, p.~3903, 2022.

\bibitem{umesh2019fully}
U.~Vazirani and T.~Vidick, ``Fully device independent quantum key distribution,'' {\em Commun. ACM}, vol.~62, p.~133, mar 2019.

\bibitem{xu2022device}
F.~Xu, Y.-Z. Zhang, Q.~Zhang, and J.-W. Pan, ``Device-independent quantum key distribution with random postselection,'' {\em Phys. Rev. Lett.}, vol.~128, p.~110506, Mar 2022.

\bibitem{bernstein2017post}
D.~J. Bernstein and T.~Lange, ``Post-quantum cryptography,'' {\em Nature}, vol.~549, no.~7671, pp.~188--194, 2017.

\bibitem{NISTFIPS203}
``Nist standard for fips203.'' \url{https://csrc.nist.gov/pubs/fips/203/final}.

\bibitem{NISTFIPS204}
``Nist standard for fips204.'' \url{https://csrc.nist.gov/pubs/fips/204/final}.

\bibitem{NISTFIPS205}
``Nist standard for fips205.'' \url{https://csrc.nist.gov/pubs/fips/205/final}.

\bibitem{bos2018crystals}
J.~Bos, L.~Ducas, E.~Kiltz, T.~Lepoint, V.~Lyubashevsky, J.~M. Schanck, P.~Schwabe, G.~Seiler, and D.~Stehl{\'e}, ``Crystals-kyber: a cca-secure module-lattice-based kem,'' in {\em 2018 IEEE European Symposium on Security and Privacy (EuroS\&P)}, pp.~353--367, IEEE, 2018.

\bibitem{lyubashevsky2020crystals}
V.~Lyubashevsky, L.~Ducas, E.~Kiltz, T.~Lepoint, P.~Schwabe, G.~Seiler, D.~Stehl{\'e}, and S.~Bai, ``Crystals-dilithium,'' {\em Algorithm Specifications and Supporting Documentation}, 2020.

\bibitem{bernstein2015sphincs}
D.~J. Bernstein, D.~Hopwood, A.~H{\"u}lsing, T.~Lange, R.~Niederhagen, L.~Papachristodoulou, M.~Schneider, P.~Schwabe, and Z.~Wilcox-O’Hearn, ``Sphincs: practical stateless hash-based signatures,'' in {\em Annual international conference on the theory and applications of cryptographic techniques}, pp.~368--397, Springer, 2015.

\bibitem{bernstein2019sphincs+}
D.~J. Bernstein, A.~H{\"u}lsing, S.~K{\"o}lbl, R.~Niederhagen, J.~Rijneveld, and P.~Schwabe, ``The sphincs+ signature framework,'' in {\em Proceedings of the 2019 ACM SIGSAC conference on computer and communications security}, pp.~2129--2146, 2019.

\bibitem{diamanti_practical_2016}
E.~Diamanti, H.-K. Lo, B.~Qi, and Z.~Yuan, ``Practical challenges in quantum key distribution,'' {\em npj Quantum Information}, vol.~2, pp.~1--12, Nov. 2016.

\bibitem{wang2021experimental}
L.-J. Wang, K.-Y. Zhang, J.-Y. Wang, J.~Cheng, Y.-H. Yang, S.-B. Tang, D.~Yan, Y.-L. Tang, Z.~Liu, Y.~Yu, {\em et~al.}, ``Experimental authentication of quantum key distribution with post-quantum cryptography,'' {\em npj quantum information}, vol.~7, no.~1, p.~67, 2021.

\bibitem{djordjevic2020joint}
I.~B. Djordjevic, ``Joint qkd-post-quantum cryptosystems,'' {\em IEEE Access}, vol.~8, pp.~154708--154712, 2020.

\bibitem{ThisWork}
{\em Practical hybrid PQC-QKD protocols with enhanced security and performance}.
\newblock Forthcoming work.

\bibitem{Simmons1990how}
G.~J. Simmons, ``How to (really) share a secret,'' in {\em Advances in Cryptology --- CRYPTO' 88}, (New York, NY), pp.~390--448, Springer New York, 1990.

\bibitem{eldar2016efficient}
In 2016, Eldar and Shor suggested an efficient quantum algorithm for the lattice problems \url{https://arxiv.org/abs/1611.06999}, but the paper was later withdrawn.

\bibitem{chen2024quantum}
Y.~Chen, ``Quantum algorithms for lattice problems.'' Cryptology {ePrint} Archive, Paper 2024/555, 2024.

\bibitem{jao2011towards}
D.~Jao and L.~De~Feo, ``Towards quantum-resistant cryptosystems from supersingular elliptic curve isogenies,'' in {\em Post-Quantum Cryptography: 4th International Workshop, PQCrypto 2011, Taipei, Taiwan, November 29--December 2, 2011. Proceedings 4}, pp.~19--34, Springer, 2011.

\bibitem{castryck2023efficient}
W.~Castryck and T.~Decru, ``An efficient key recovery attack on sidh,'' in {\em Annual International Conference on the Theory and Applications of Cryptographic Techniques}, pp.~423--447, Springer, 2023.

\bibitem{huang2024vacuum}
Y.~Huang, F.~Salces-Carcoba, R.~X. Adhikari, A.~H. Safavi-Naeini, and L.~Jiang, ``Vacuum beam guide for large scale quantum networks,'' {\em Physical Review Letters}, vol.~133, no.~2, p.~020801, 2024.

\bibitem{liu2022quantum}
J.~Liu, C.~T. Hann, and L.~Jiang, ``Data centers with quantum random access memory and quantum networks,'' {\em Physical Review A}, vol.~108, no.~3, p.~032610, 2023.

\bibitem{liu2024quantum}
J.~Liu and L.~Jiang, ``{Quantum Data Center: Perspectives},'' {\em IEEE Network}, vol.~38, pp.~160--166, 2024.

\bibitem{kimble2008quantum}
H.~J. Kimble, ``The quantum internet,'' {\em Nature}, vol.~453, no.~7198, pp.~1023--1030, 2008.

\end{thebibliography}

\end{document}